# Oakridge PPU Magnets: Results and Measurements

J. DiMarco, D. Harding, V. Kashikhin, O. Kiemschies, M. Kifarkis, A. Makulski,
J. Nogiec, S. Stoynev, T. Strauss, M. Tartaglia, P. Thompson

*Abstract*— The Spallation Neutron Source (SNS) at Oak Ridge National Laboratory (ORNL) is being upgraded from 1.0 GeV to 1.3 GeV (or 1.4 to 2.8 MW). Several water-cooled magnets have been upgraded to transport 30% higher beam energy. Fermilab contributed the magnet design for the new chicane magnets and injection/extraction septum. Designing the magnets was a challenging task because the new magnets required good combined integrated field quality and needed to occupy the old magnets space but with about 20% greater integrated magnetic field. Additional strong requirements applied to the magnets fringe field so as not to disturb the circulating beam. After fabrication of the magnets, an extensive measurement campaign was developed and performed at Fermilab's Magnet Test Facility. The measurements needed to assess magnet performance and provide comparison to design calculations. These included verification of field strength and harmonics along an 8 m length and 200 mm good field diameter for the chicane dipoles, end-field Hall probe mapping of these magnets, and measurements along two differently curved trajectories within the ~3 m septum gradient magnet. Details of the measurements and systems are presented along with results and comparison to field models.

*Index Terms*— Accelerator magnets, electromagnetic measurements, magnetic measurements, magnet alignment.

## I. INTRODUCTION

The Proton Power Upgrade (PPU) Project at the Spallation Neutron Source at Oak Ridge National Laboratory increased the proton beam power capability from 1.4 to 2.8 MW [1]. Upon completion of the project, 2 MW of beam power are available for neutron production at the existing first target station with the remaining beam power available for the future second target station [2]. Partnering in this work, Fermilab designed, fabricated and tested chicane dipole magnets, OCA001 ("D2, OCA") and OCB001 ("D3, OCB"), a chicane magnet spare, and also a combined function 'dump septum' magnet. The designs of these are described in detail in [3]. This paper reports on the magnet measurements developed and performed to verify magnet performance and to compare with design. Field strength and field quality, and measurements in fringe and low-field regions are presented.

## II. CHICANE DIPOLES

For the PPU upgrade, two chicane magnets were needed to provide the beam bending in the horizontal plane, with a stripping foil placed in the region between the magnets. The field of the chicanes is fairly low, less than 0.25 T, and they have an aperture of ~240 mm [3]. The innovative design that was developed features different length iron for the top/bottom magnet poles (1.31 m / 0.91 m respectively for D3), and these are reversed for the second magnet of the pair (D2) by just inverting the magnet – thus allowing one design to be used for both. The magnet cross-section is shown in Fig. 1 and magnets under test are shown in Fig. 2.

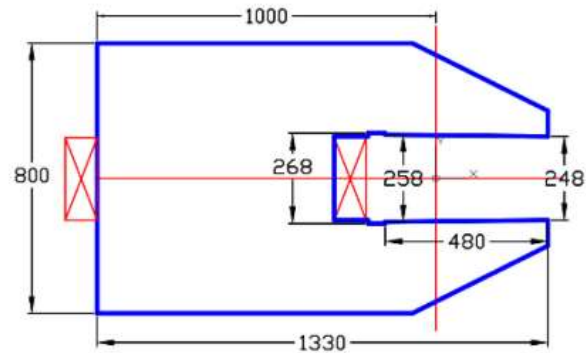

Fig. 1 Chicane Magnet cross section

A steel shield is necessary to attenuate the magnetic field upstream of a second stripping foil located downstream of the D3 dipole. The actual PPU shield material was shipped from Oak Ridge to the test facility, where it was assembled into an aluminum support structure that was mounted at the appropriate location next to the OCB magnet. Due to the rather large (~2 ton) attractive magnetic force on the shield, measurements of the structure deflection versus current were made to ensure safe operation; essentially no deflection was detected. The good field region requires harmonics less than 0.1% at a reference radius of 100 mm, and the full field integral for the two-magnet system requires a at least a 7 m length probe because of the large fringe fields. The magnets have steel end shims for tuning of the field harmonics.

### A. Integral strength and harmonics measurements

For 1.3 GeV operation, the OCA/OCB are powered with 1711 A/ 1278 A respectively, and for 1.0 Gev, 1427 A/ 1066 A. The OCA and OCB magnets were tested when powered individually and also together. The integrated field strength and field quality were measured with a Single Stretched Wire (SSW) system [4] at the 1.0 and 1.3 GeV currents. The Single Stretched Wire (SSW) measurement system is comprised of a pair of portable, high-accuracy XY-motion stages with 1 micron accuracy and sub-micron resolution, between which a single wire is attached and moved to measure magnetic field strength (the return loop for the wire lies stationary at the bottom of the magnet). Berylium Copper wire (0.1 mm diameter) was used near its maximum tension in order to minimize sag. Measurements were made with an 8m wire length. The stages have +/- 75 mm travel horizontally and vertically from their center.

[1] Submitted for review August 4, 2025. This work was supported by the U.S. Department of Energy, Office of Science, Office of High Energy Physics. The authors are with Fermi National Accelerator Laboratory, P.O. Box 500, Batavia, IL 60510, USA, (corresponding author e-mail: dimarco@fnal.gov).

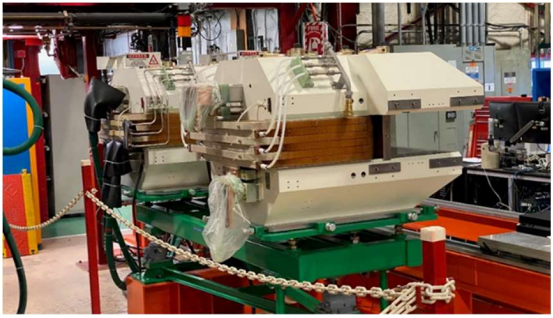

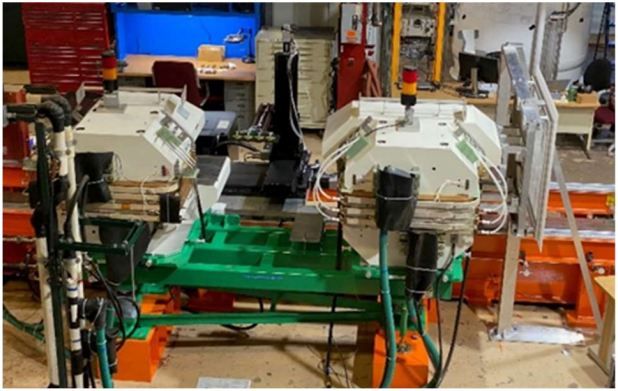

Fig 2. Chicane Magnet Pair under Test at Fermilab

Measurements were taken on the perimeter of a circle, moving to each of 32 discrete vertex positions at a radius of 65 mm and then executing +/- 10 mm XY moves to measure flux. The resulting analysis of these data provides both integrated field strength and harmonics. In order to improve measurement sensitivity for reporting at 100 mm reference radius, a square wire perimeter with 28 positions was also used – which allows the wire, when in the corner positions, to reach to 75 mm in either the X or Y coordinate, and 65 mm in the other. The radial reach for these positions is thus ~99 mm. The complex field, $B$, is expressed as

$$B(z) = B_y + i * B_x = \sum_{n=1}^{\infty} (B_n + iA_n)\left(\frac{z}{R}\right)^{n-1}$$

where $B_n$, $A_n$ are the harmonic field coefficients, $z = x + iy$ is the (complex) position from magnet center in the aperture, and $R$ is the reference radius (100 mm). Defining a sensitivity, $K_n$, for each vertex $v$, with wire XY motion steps of +/- $d$, $L=1$ (since integral measurement) and harmonic order, $n$, as

$$K_n(v) = \frac{LR}{n} * \left(\left(\frac{d^+}{R}\right)^{n-1} - \left(\frac{d^-}{R}\right)^{n-1}\right)$$

the harmonic coefficients, $C_n = B_n + iA_n$, can be determined from the fluxes of the XY measurements at each vertex, $\varphi_v$, and the matrix expression

$$\begin{pmatrix} (K_1(v_{1x})) & (K_2(v_{1x})) & (K_3(v_{1x})) & \cdots & (K_{14}(v_{1x})) \\ (K_1(v_{1y})) & (K_2(v_{1y})) & (K_3(v_{1y})) & & (K_{14}(v_{1y})) \\ (K_1(v_{2x})) & (K_2(v_{2x})) & (K_3(v_{2x})) & & (K_{14}(v_{2x})) \\ (K_1(v_{2y})) & (K_2(v_{2y})) & (K_3(v_{2y})) & \cdots & (K_{14}(v_{2y})) \\ & & \vdots & & \\ (K_1(v_{28y})) & (K_2(v_{28y})) & (K_3(v_{28y})) & \cdots & (K_{14}(v_{28y})) \end{pmatrix} \begin{pmatrix} C_1 \\ C_2 \\ C_3 \\ \vdots \\ \vdots \\ C_{14} \end{pmatrix} = \begin{pmatrix} \varphi_{1x} \\ \varphi_{1y} \\ \varphi_{2x} \\ \vdots \\ \vdots \\ \varphi_{28y} \end{pmatrix}$$

The field amplitudes of the measurements with circular perimeter are shown in Fig. 3 and those with the square perimeter in Fig. 4. Since a linear decrease in the logarithm of harmonic amplitude for unallowed orders is typically expected, the observed increase in high order harmonics for the circular perimeter indicates insufficient sensitivity; whereas the results of the square roughly flatten out (< 10 units) above n=6, indicating adequate sensitivity at least up to this order. The low-order multipoles measured for each of the magnets at 1.3 GeV currents, and with both powered, are summarized in Table 1, showing that the large skew quadrupole harmonics from the asymmetric poles cancel when paired. A comparison of the 2-magnet result with the calculated harmonics from design is shown in Figure 5, to very good agreement. Various changes to the shims were tried to see if the octupole could be further reduced, but the original configuration of shims (Fig. 5) was shown to give the lowest overall harmonics. The results meet the specification of having harmonics less than 10 units (1e−3).

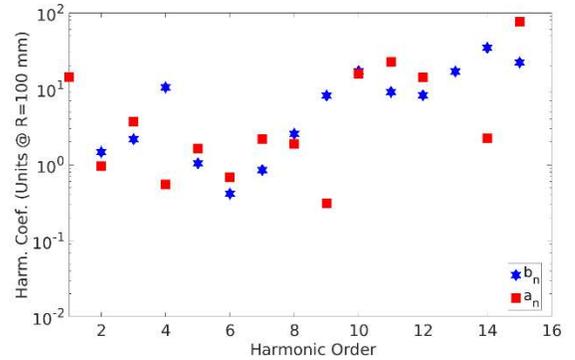

Fig 3. Harmonic coefficients of the measurements with circular perimeter (1 unit is 1e-4 of the main field at R).

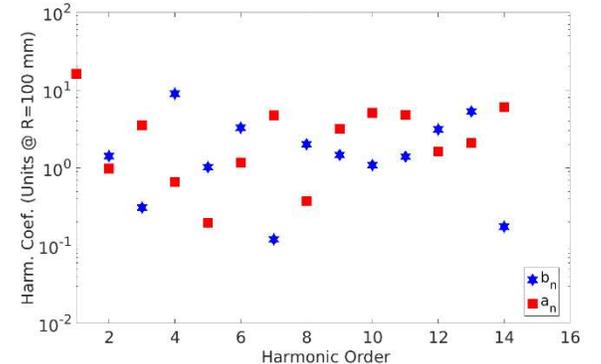

Fig 4. Harmonic coefficients of the measurements with circular perimeter (1 unit is 1e-4 of the main field at R).

The SSW measured integral strength is 0.5975 T-m, whereas the calculated strength from the model gives a value of 0.60273 T-m, indicating that the model is high by about 87 units (0.87%) compared to the measured value. This is consistent with the lower measured strength values observed in the Hall probe results of about 1% when compared to the model. The reason for this discrepancy was not fully understood, but the measured results of the integrated field strength are more than sufficiently accurate to obtain required fields for operation at both 1.3 and 1.0 GeV beam energies.

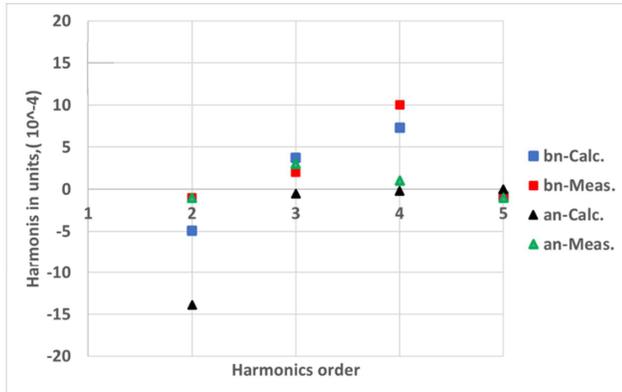

Fig 5. Comparison of the 2-magnet harmonics measurement result with the calculation from model.

TABLE I
FIELD QUALITY OF CHICANE MAGNETS

| Field Quality Summary | 1.0 Gev OCA+OCB | 1.3 Gev OCA (1711 A) | 1.3 Gev OCB (1277 A) | 1.3 Gev OCA+OCB |
|---|---|---|---|---|
| int. str. (Tm) | 0.4999 | 0.3441 | 0.2532 | 0.5975 |
| harm. Coef (units) | | | | |
| b2 | -2 | -6 | -1 | -1 |
| a2 | -3 | 59 | -87 | -1 |
| b3 | 0 | -2 | 5 | 0 |
| a3 | 3 | 11 | -8 | 4 |
| b4 | 9 | 7 | 11 | 9 |
| a4 | -1 | 2 | -4 | -1 |
| b5 | 2 | 1 | 1 | 1 |
| a5 | -2 | 0 | -2 | 0 |

*B. Hall probe studies*

An array of Hall probes was used to map the magnetic field within, between, and beyond the OCA and OCB chicane dipoles, with specialized study in particular regions. CERN/NIKHEF Hall probes [5] and an electronic readout system were used for the bulk of "point scan" magnetic field measurements [6,7]. These probes have three Hall elements mounted on the face of a small glass cube (~5mm on a side) to provide a 3-D field measurement within a small volume (Fig 6). The packaged element has a G-10 support with 3 metal spheres for mounting to a support structure. Field measurements are digitized by an on-board 24-bit ADC, and the probes are calibrated by CERN to a precision of one part in 10,000 with spherical harmonic coefficients that correct for small non-orthogonality of the Hall elements. Each probe has a unique address, and multiple probes are daisy-chained on an SPI bus that is interrogated by a readout module, the mBatcan, which communicates digital information over CANBUS to a PC (via a CANBUS-USB interface). The probe motions, current control and readouts are scripted in textual configuration files that control a LabView program (EMMA, Enhanced Magnetic Measurement Applications) developed by the T&I Software and Computing group [8]. The Hall array is shown in Figs. 7 and 8. The field measured through the 2-magnet volume is shown in Fig. 9.

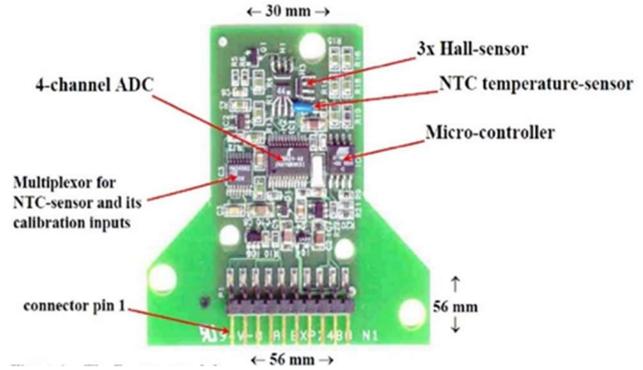

Fig 6 CERN/NIKHEF 3D Hall probe

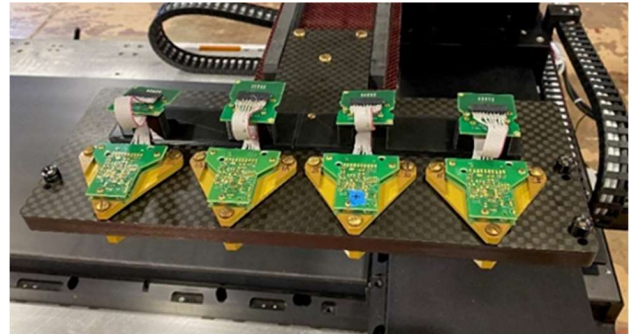

Fig 7. Hall probe array for the field measurements

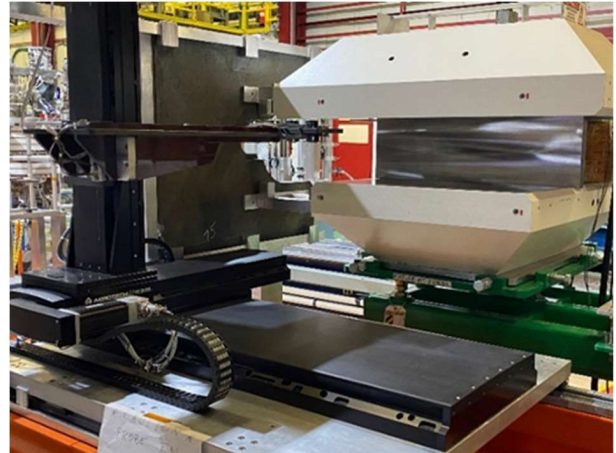

Fig 8. Hall probe measurement in the 2-magnet volume

The field maps in Fig. 9 show very good agreement with model predictions throughout the large magnetic volume, with some small deviations that may be related to actual, versus assumed, magnet steel properties, or possibly due to the steel support structure. The foil region maps provide the very detailed and high-quality information needed for proper placement of the D2 stripping foil (Fig. 10). The field strength in the second foil region beyond the magnetic shield (at an axial distance of 102.6 cm from the center of the D3 dipole) also needed to be below 100 G. Measurements with a 3-axis Metrolab Hall probe found 80.3 G at a distance of 102 cm, satisfying this requirement as well.

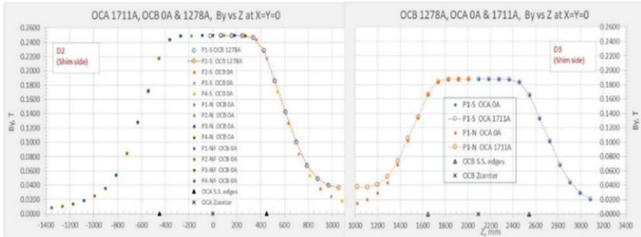

Fig 9. Hall probe field map

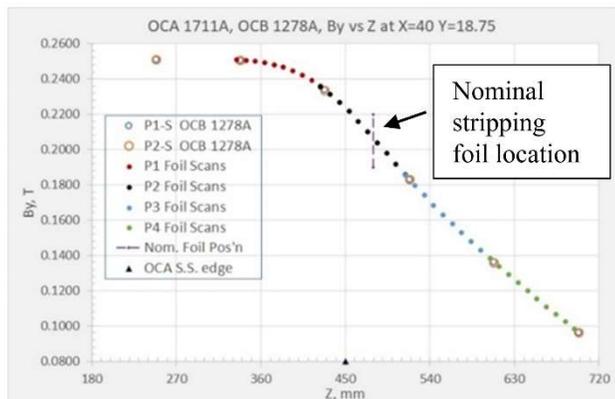

Fig 10. Hall probe measurement detail for the verification of the striping foil location

To verify the field strength and accuracy of the current and Hall probe readings, measurements were taken using a Metrolab PT-2026 NMR instrument and probe at the "measurement center" (centered in Y and Z, and 150 mm from the Stainless-Steel plate in X). The readings were very reproducible and changed by only 3 G (from 0.2493 to 0.2496 T) after returning from a 1900A pre-ramp. The model prediction is 0.2500 T.

### III. INJECTION DUMP SEPTUM MAGNET

The injection dump septum magnet (OSD001), shown in Fig. 11, provides simultaneous bending of 1.3 GeV H− and H0 (proton) beams. This requires a magnetic field with a strong quadrupole term, and a large 150 mm beam separation, with different bending angles, for H− and H0 [3]. The quadrupole field component in the magnet gap was obtained from a hyperbolic pole tip profile. In addition, 15° chamfers on both pole ends were incorporated to provide a shorter effective pole length for H− than for H0 [3].

Stretched wire measurements on OSD001 were performed with the magnet at 4582.4 A. The SSW stages were surveyed so that the wire was defined to be centered when passing through [x,z] points which were centered both on the entrance and exit points of the two curved trajectories. Placement was done to ~1mm. Measurements were made with the wire at this position, and with the wire centered on the $H^0$ and $H^-$ path chords (i.e., the x start position was changed by step sizes of 0.06 m at the stage near the z=0 end, and by step sizes of 0.05 m at the stage, near the z=-2.5 m end). These latter two are compared to calculations. At each of these start positions, the wire was moved in 3 mm steps to determine the magnet integral dipole strength. Results are shown in Table 2, with calculations performed using the measured B(H) curve of the iron. Agreement is at the level of 0.1%, with uncertainty in the measured integrals being about 0.05%. Note that 'Enter' and 'Exit' coordinates in the table are the actual stage positions, but that the wire passes through the points indicated above.

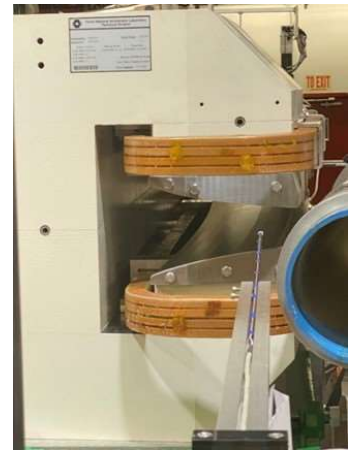

Fig 11. Hall probe measurement of the OSD001 magnet

Table II

| SSW Measurements vs Simulation | | | Simulated | Measured | |
|---|---|---|---|---|---|
| Trajectory chord | Enter [x,z] (m) | Exit [x,z] (m) | Integral By, (T-m) | Integral By, (T-m) | % diff |
| $H^0$ | [0.475, 0.655] | [0.179, -2.724] | 1.0992 | 1.0980 | -0.11 |
| $H^-$ | [0.355, 0.655] | [0.079, -2.724] | 0.8030 | 0.8021 | -0.11 |

The Single Stretched Wire system was also used to measure the integral field in the shielded septum area. The wire was placed at the midplane of the field free region near the inner wall closest to the magnet, and measurements performed with both +/- 3 mm steps and +/- 10 mm wire steps across the aperture, moving from "0" (~5 mm from wall) to about 130 mm away from the magnet (Fig. 12). The measurement results are plotted in Fig. 13 and indicate a maximum integrated field near the inner wall of about 17 G-m, dropping to about 8 G-m when 130 mm away. The calculations show about 28 G-m dropping to 14 G-m at these two positions. Note that the simulation uses a solid shield, whereas the actual shield is made of concentric pipes, which should be more efficient at shielding; therefore, the measured field being lower than simulation would also seem consistent with expectation. Both measured and simulated results showed a drop of about half the maximum field moving from the inner position to 130 mm away.

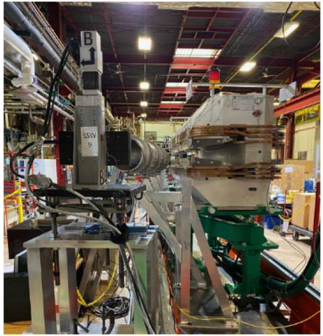

Fig 12. SSW measurement in the shielded septum area

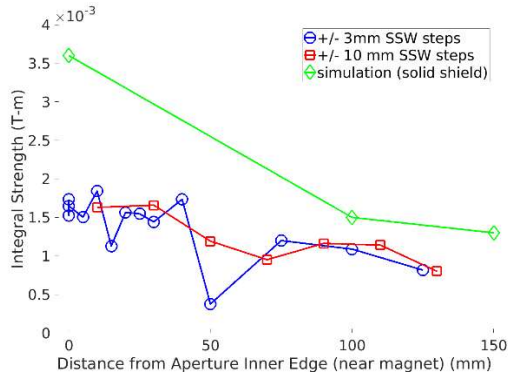

Fig 13. Field strength inside the shielded septum area

In order to measure directly along the nominal H$^-$ and H$^0$ trajectories of the dump magnet, a single wire-bundle flatcoil was fabricated (Fig. 14). The wire bundle consists of 12 insulated wires twisted together but connected in series (so as to multiply the signal), within a jacket of 2.4 mm outer diameter. Two such wire bundles were laid into grooves which followed the H$^-$ and H$^0$ beam trajectories, with the return loop of each wire bundle lying stationary along the back-leg of the magnet (Fig. 14). The grooves are machined into a 3.35 m-long (132"-long) carbon fiber plate, which is supported by the stages at its ends, and by low-friction roller assemblies within the magnet. The change in flux induced in each wire bundle loop stems from the precise translation of the flatcoil in the magnetic field. Since this motion is controlled using the SSW stages, which have 1 micron motion accuracy, a measurement of the integrated voltage during movement yields a high-accuracy measurement of the dipole and gradient field integrals on these paths. Table 3 compares the simulation with the measured integral results for flatcoil motion step of +/- 3mm. Agreement is better than about 0.15% for both H$^-$ and H$^0$; these have values of 1.2371 T-m and 0.8527 T-m respectively.

Table III

| Flatcoil Trajectories | Simulated Integral By, (T-m) | Measured Integral By, (T-m) | s.d. (T-m) | Meas-Siml % diff |
|---|---|---|---|---|
| H0 | 1.2390 | 1.2371 | 0.0003 | -0.15 |
| Hmin | 0.8528 | 0.8527 | 0.0003 | -0.01 |

Hall probe scans of the dump septum magnet were performed using a SENIS 3-axis Analog Magnetic Field Transducer (model F3A-03A05F-A02T2K5J, shown in Fig 11.

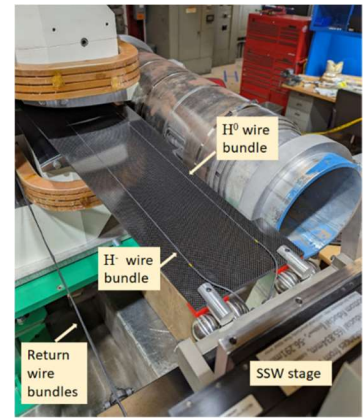

Fig 14. Curved single wire bundle flat coil

This probe has a +/-2 T range that is calibrated to 0.1% by the vendor. It has a small profile, essentially an integrated circuit package with 5 mm width and 2 mm height, and a well-defined Field Sensitive Point of less than 0.15 mm dimensions. The analog readout electronics compensate for the Planar Hall effect, and provide voltages corresponding to probe temperature, Bx, By, and Bz with a 5 V/T transfer function and 2.5 kHz bandwidth. Those voltages were digitized using a Keithley 2700 Multiplexing Digital Multimeter and recorded by the EMMA [8] measurement system along with the probe motion stage coordinates and magnet current. Sample data showing field profile across the aperture are shown in Fig. 15.

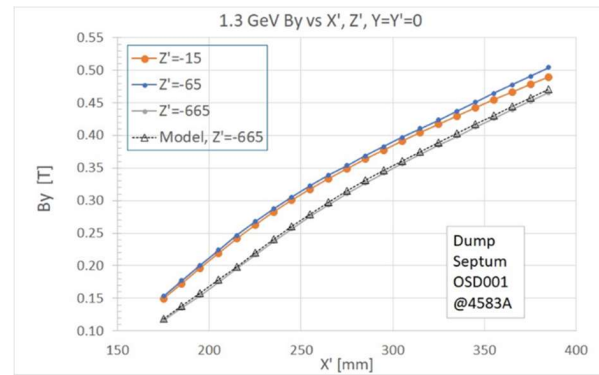

Fig 15. Hall Probe Measurement of the OSD001 magnet

V. CONCLUSION

A comprehensive magnetic measurement program was completed for the PPU Chicane Dipoles and Injection Dump Septum magnets. Comparisons validating the magnetic model were obtained for the chicane dipole volume and fringe and stripping foil regions with the Hall probe system, and integral field strength and harmonic field quality were obtained using the SSW system. The dump septum had both straight wire integrals and curved beam trajectory measurements that also confirmed model predictions, including in the field free region. Results indicated that the magnets successfully met ORNL PPU requirements, as has now been demonstrated [9][10].